\documentclass[letterpaper, preprint, paper,11pt]{AAS}	

\usepackage{bm}
\usepackage{amsmath}
\usepackage[colorlinks=true, pdfstartview=FitV, linkcolor=black, citecolor= black, urlcolor= black]{hyperref}
\usepackage{overcite}
\usepackage{footnpag}			      	

\usepackage[utf8]{inputenc}
\usepackage{textcomp}
\usepackage{multicol}
\usepackage{lscape}

\usepackage{cleveref}
\usepackage{algorithm}
\usepackage{algpseudocode}
\usepackage{graphicx}
\usepackage{amssymb}
\usepackage[version=4]{mhchem}
\usepackage{siunitx}
\usepackage{longtable,tabularx}
\usepackage{xcolor} 
\usepackage{wrapfig} 
\usepackage{subcaption} 
\usepackage{array}

\title{Run Time Assurance for \\ Autonomous Spacecraft Inspection\thanks{\NoCaseChange{Approved for Public Release, Case Numbers  AFRL-2023-0599. This work was supported by the Air Force Research Laboratory (AFRL) Safe, Trusted Autonomy for Responsible Spacecraft (STARS) Seedlings for Disruptive Capabilities Program (SDCP). The views expressed are those of the authors and do not reflect the official guidance or position of the United States Government, the Department of Defense or of the United States Air Force.}}}

\author{Kyle Dunlap\thanks{AI Scientist, RDT\&E Division, Parallax Advanced Research, 4035 Colonel Glenn Hwy, Beavercreek, OH, 45431.},  
David van Wijk\thanks{Graduate Research Fellow, Aerospace Engineering, Texas A\&M University, Land, Air, and Space Robotics (LASR) Laboratory, 1188 Nuclear Science Rd, College Station, TX, 77845.},
\ and Kerianne L. Hobbs\thanks{Safe Autonomy Lead, Autonomy Capability Team 3, Air Force Research Laboratory, 2241 Avionics Circle, Wright-Patterson AFB, OH, 45433.
}
}

\PaperNumber{23-113}

\begin{document}

\maketitle

\begin{abstract}
As autonomous systems become more prevalent in the real world, it is critical to ensure they operate safely. One approach is the use of Run Time Assurance (RTA), which is a real-time safety assurance technique that monitors a primary controller and intervenes to assure safety when necessary. As these autonomous systems become more complex, RTA is useful because it can be designed completely independent of the primary controller, thus decoupling mission completion and safety assurance. This paper develops several translational motion safety constraints for a multi-agent autonomous spacecraft inspection problem, where all of these constraints can be enforced with RTA. A comparison is made between centralized and decentralized control, where simulations of the inspection problem then demonstrate that RTA can assure safety of all constraints. Monte Carlo analysis is then used to show that no scenarios were found where the centralized RTA cannot assure safety. While some scenarios were found where decentralized RTA cannot assure safety, solutions are discussed to mitigate these failures. 

\end{abstract}


\section{Introduction}

In order for autonomy to be used safely on mission-critical systems, assurance methods must be considered. For space systems in particular, a mistake or fault could result in damage to multi-million or multi-billion dollar equipment, as well as potential loss of invaluable data and space-based services. One method of assuring safety is with the use of Run Time Assurance (RTA) \cite{schierman2020runtime}, which is an online safety assurance technique designed to filter potentially unsafe inputs from a primary controller and intervene to assure safety of the system when necessary. RTA systems decouple the task of safety assurance from all other control objectives, allowing it to scale well to increasingly complex systems. One type of RTA filter is the Active Set Invariance Filter (ASIF) \cite{gurriet2018online}, which utilizes Control Barrier Functions (CBF) \cite{ames2019control} to minimize deviation from the primary controller while still assuring safety of the system. Additionally, the ASIF approach allows multiple constraints to be enforced at once, without needing complex switching or decision logic.

\begin{figure}[htb!]
    \centering
    \includegraphics[width=.5\textwidth]{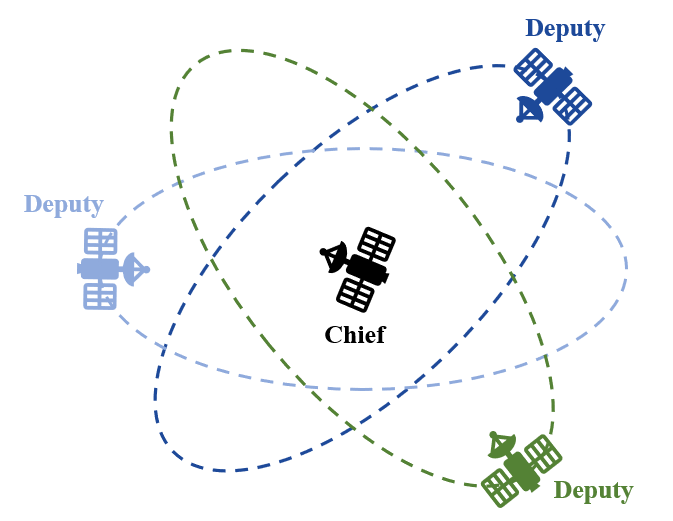}
    \caption{Inspection Problem.}
    \label{fig:Inspection}
\end{figure}

This paper focuses on assuring safety for an autonomous spacecraft inspection problem, where multiple active deputy spacecraft cooperate to inspect a passive, stationary chief spacecraft. A group of deputies inspecting a chief is shown in Figure \ref{fig:Inspection}. Several safety constraints on position and velocity are developed and enforced with RTA. While most of these constraints can be enforced through ASIF RTA, a specific case is shown where alternate RTA methods would better assure safety. 

CBFs for online safety assurance are becoming increasingly popular, with applications to autonomous driving \cite{ames2016control}, segways \cite{taylor2020learning}, fixed-wing aircraft \cite{squires2022composition}, and many more. For spacecraft related problems, CBFs have been used to assure safety for docking \cite{breeden2021guaranteed}, rendezvous \cite{agrawal2021safe}, proximity operations \cite{mote2021natural}, and attitude control \cite{wu2021attitude}.
The safety constraints in this work are adapted to an ASIF RTA approach from safety requirements for automatic spacecraft maneuvering. \cite{hobbs2021risk, hobbs2020elicitation}. This work builds upon previous constraints developed for spacecraft docking \cite{dunlap2021comparing} and inspection \cite{hibbard2022guaranteeing}.

The main contributions of this paper are developing mathematical constraints for the spacecraft inspection problem, comparing centralized and decentralized RTA methods, performing Monte Carlo analyses to show that all constraints can be enforced simultaneously, and considering the case where ASIF RTA can be combined with a switching-based RTA approach.

\section{Run Time Assurance}


This paper considers control affine dynamical systems, where a continuous-time system model is given by a system of ordinary differential equations,
\begin{equation} \label{eq:fxgu}
   \boldsymbol{\dot{x}} = f(\boldsymbol{x}) + g(\boldsymbol{x})\boldsymbol{u}.
\end{equation}
Here, $\boldsymbol{x} \in \mathcal{X} \subseteq \mathbb{R}^n$ denotes the state vector, $\boldsymbol{u}\in \mathcal{U} \subseteq\mathbb{R}^m$ denotes the control vector, and $f:\mathcal{X} \rightarrow \mathbb{R}^n$ and $g:\mathcal{X} \rightarrow \mathbb{R}^{n \times m}$ are locally Lipschitz continuous functions. Additionally, $\mathcal{X}$ is the set of all possible state values and $\mathcal{U}$ is the admissible control set.

\begin{figure}[htb!]
    \centering
    \includegraphics[width=.8\textwidth]{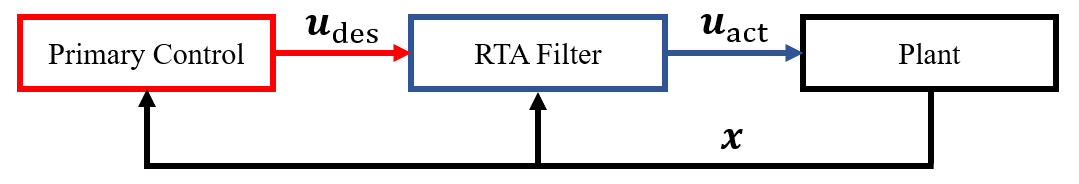}
    \caption{Feedback control system with RTA.}
    \label{fig:RTA_Filter}
\end{figure}

As shown in Figure \ref{fig:RTA_Filter}, a feedback control system with RTA is divided into a performance-focused primary controller and a safety-focused RTA filter. The primary controller first computes a desired control input, $\boldsymbol{u}_{\rm des}$, based on the state $\boldsymbol{x}$. The RTA filter evaluates $\boldsymbol{u}_{\rm des}$ at the current state and modifies it as necessary to produce a safe control input, $\boldsymbol{u}_{\rm act}$, which is then passed to the plant. In Figure \ref{fig:RTA_Filter}, the primary controller is highlighted red to indicate low safety confidence, while the RTA filter is highlighted blue to indicate high safety confidence. This structure allows the designer to isolate unverified or unsafe components of the control system.

For a dynamical system, safety can be defined by a set of $M$ inequality constraints $\varphi_i(\boldsymbol{x}): \mathcal{X} \to \mathbb{R}$,  $\forall i \in \{1,...,M\}$, where $\varphi_i(\boldsymbol{x}) \geq 0$ when the constraint is satisfied. The set of states that satisfies all of these constraints is known as the \textit{allowable set} $\mathcal{C}_{\rm A}$, and is defined as,
\begin{equation}
    \mathcal{C}_{\rm A} := \{\boldsymbol{x} \in \mathcal{X} \, | \, \varphi_i(\boldsymbol{x}) \geq 0, \forall i \in \{1,...,M\} \}.
\end{equation}
A state $\boldsymbol{x}(t_0)$ is then said to be safe if it lies in a \textit{forward invariant} subset of $\mathcal{C}_{\rm A}$ known as the \textit{safe set} $\mathcal{C}_{\rm S}$, where,
\begin{equation}
    \boldsymbol{x}(t_0) \in \mathcal{C}_{\rm S} \Longrightarrow \boldsymbol{x}(t) \in \mathcal{C}_{\rm S}, \forall t \geq t_0.
\end{equation}
Note that the control input $\boldsymbol{u}$ is bounded by the admissible control set $\mathcal{U}$, and therefore $\mathcal{C}_{\rm S}$ is highly dependent on the controller. $\mathcal{C}_{\rm S}$ is then said to be \textit{control invariant} if there exists a control law $\boldsymbol{u} \in \mathcal{U}$ that renders $\mathcal{C}_{\rm S}$ forward invariant. $\mathcal{C}_{\rm S}$ can be defined explicitly by a set of $M$ control invariant inequality constraints, $h_i(\boldsymbol{x}): \mathcal{X}\to \mathbb{R}$,  $\forall i \in \{1,...,M\}$, where again $h_i(\boldsymbol{x}) \geq 0$ when the constraint is satisfied. $\mathcal{C}_{\rm S}$ is then defined as,
\begin{equation} \label{eq: explicitly_defined_safe_set}
    \mathcal{C}_{\rm S} := \{\boldsymbol{x}\in\mathcal{X} \, | \, h_i(\boldsymbol{x})\geq 0, \forall i \in \{1,...,M\} \}.
\end{equation}

While there are many approaches to developing an RTA filter,\cite{hobbs2021run,dunlap2022run} ASIF is an optimization-based algorithm designed to be minimally invasive towards the primary controller while still respecting multiple safety constraints. The ASIF algorithm uses CBFs to define safety, and a Quadratic Program (QP) to compute a safe control input. The ASIF algorithm used in this paper is defined as follows.

\noindent \rule{1\columnwidth}{0.7pt}
\noindent \textbf{Active Set Invariance Filter}
\begin{equation}
\begin{gathered}
\boldsymbol{u}_{\rm act}(\boldsymbol{x}, \boldsymbol{u}_{\rm des})= \underset{\boldsymbol{u} \in \mathcal{U}}{\text{argmin}} \left\Vert \boldsymbol{u}_{\rm des}-\boldsymbol{u}\right\Vert_2 ^{2}\\
\text{s.t.} \quad BC_i(\boldsymbol{x},\boldsymbol{u})\geq 0, \quad \forall i \in \{1,...,M\}
\end{gathered}\label{eq:optimization}
\end{equation}
\noindent \rule[7pt]{1\columnwidth}{0.7pt}

Here, $BC$ represents one of $M$ barrier constraints, which are satisfied when $BC_i(\boldsymbol{x},\boldsymbol{u})\geq 0$. The objective of these barrier constraints is to render $\mathcal{C}_{\rm S}$ forward invariant under $\boldsymbol{u}_{\rm act}$. Nagumo's condition \cite{nagumo1942lage} is used to do this, where the boundary of $\mathcal{C}_{\rm S}$ is examined to ensure $\dot{h}_i(\boldsymbol{x}) \geq 0$, thus causing $\boldsymbol{x}$ to never leave $\mathcal{C}_{\rm S}$. This is written as,
\begin{equation}
    \dot{h}_i(\boldsymbol{x}) = \nabla h(\boldsymbol{x}) \dot{\boldsymbol{x}} = L_f h_i(\boldsymbol{x}) + L_g h_i (\boldsymbol{x}) \boldsymbol{u} \geq 0,
\end{equation}
where $L_f$ and $L_g$ are Lie derivatives of $h_i$ along $f$ and $g$ respectively. Note that this condition should only be enforced along the boundary of $\mathcal{C}_{\rm S}$, and therefore a strengthening function $\alpha(x):\mathbb{R} \rightarrow \mathbb{R}$ is introduced to relax the constraint away from the boundary. $\alpha(x)$ must be a continuous, strictly increasing class $\kappa$ function and have the condition $\alpha(0)=0$. The barrier constraint is therefore defined as,
\begin{equation} \label{eq:BC}
    BC(\boldsymbol{x},\boldsymbol{u}) := \nabla h(\boldsymbol{x}) (f(\boldsymbol{x}) + g(\boldsymbol{x})\boldsymbol{u}) + \alpha(h(\boldsymbol{x})).
\end{equation}

\section{Spacecraft Inspection Problem}

\begin{figure}[htb!]
    \centering
    \includegraphics[width=.6\textwidth]{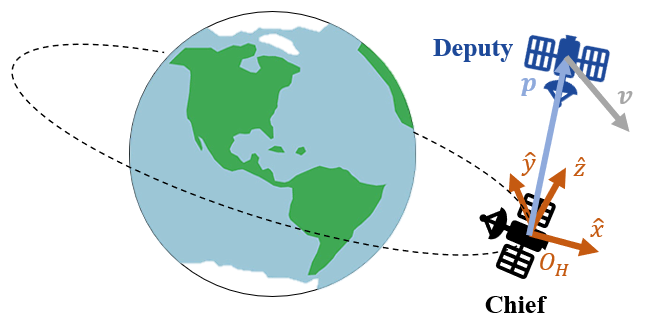}
    \caption{Hill's Frame.}
    \label{fig:Hills}
\end{figure}

This paper focuses on the task of spacecraft inspection, where multiple active deputy spacecraft are inspecting a passive chief spacecraft. The analysis takes place in Hill's frame \cite{hill1878researches}, as shown in Figure \ref{fig:Hills}. The origin of the frame, $
\mathcal{O}_H$, is located at the center of mass of the chief, the unit vector $\hat{x}$ points away from the center of the Earth, $\hat{y}$ points in the direction of motion of the chief, and $\hat{z}$ is normal to $\hat{x}$ and $\hat{y}$. The linearized relative motion dynamics between a deputy and the chief are given by the Clohessy-Wiltshire equations \cite{clohessy1960terminal}, 
\begin{equation} \label{eq: system dynamics}
    \dot{\boldsymbol{x}} = A {\boldsymbol{x}} + B\boldsymbol{u},
\end{equation}
where the state $\boldsymbol{x}=[x,y,z,\dot{x},\dot{y},\dot{z}]^T \in \mathcal{X}=\mathbb{R}^6$, the control $\boldsymbol{u}= [F_x,F_y,F_z]^T \in \mathcal{U} = [-u_{\rm max},u_{\rm max}]^3$, and
\begin{align}
\centering
    A = 
\begin{bmatrix} 
0 & 0 & 0 & 1 & 0 & 0 \\
0 & 0 & 0 & 0 & 1 & 0 \\
0 & 0 & 0 & 0 & 0 & 1 \\
3n^2 & 0 & 0 & 0 & 2n & 0 \\
0 & 0 & 0 & -2n & 0 & 0 \\
0 & 0 & -n^2 & 0 & 0 & 0 \\
\end{bmatrix}, 
    B = 
\begin{bmatrix} 
 0 & 0 & 0 \\
 0 & 0 & 0 \\
 0 & 0 & 0 \\
\frac{1}{m} & 0 & 0 \\
0 & \frac{1}{m} & 0 \\
0 & 0 & \frac{1}{m} \\
\end{bmatrix}.
\end{align}
Here, $m$ is the mass of the deputy and $n$ is the mean motion of the chief's orbit. Each deputy is independent of all others, where all follow the same dynamics.

\begin{figure}[htb!]
    \centering
    \includegraphics[width=.4\textwidth]{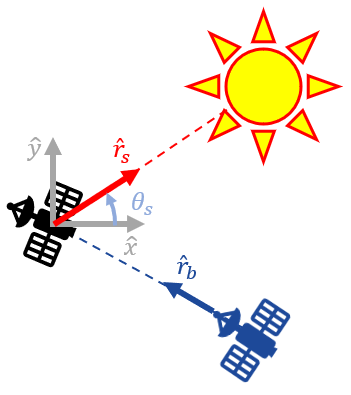}
    \caption{Sensor boresight and Sun vectors.}
    \label{fig:rb_rs}
\end{figure}

For this analysis, the attitude of each deputy is not modeled, and therefore it is assumed that each deputy is always pointing a sensor towards the chief. Letting $\boldsymbol{p}$ define the position $[x,y,z]^T$ of the deputy in Hill's frame, the unit vector defining the orientation of the deputy's sensor boresight, $\hat{r}_{b}$, is given by,
\begin{equation}
    \hat{r}_{b}=-\frac{\boldsymbol{p}}{\Vert \boldsymbol{p} \Vert_2}.
\end{equation}
Since the direction of Earth's center and chief spacecraft location are fixed in Hill's reference frame, the Sun is considered to rotate around the spacecraft. For this analysis, it is assumed that the Sun rotates at a constant rate in the $\hat{x}-\hat{y}$ plane. The unit vector pointing to the Sun, $\hat{r}_{s}$, is defined as,
\begin{equation}
    \hat{r}_{s} = [\cos{\theta_s}, \sin{\theta_s}, 0],
\end{equation}
where $\theta_s$ is the angle of the Sun with respect to the $\hat{x}$-axis, and $\dot{\theta}_s=-n$. $\hat{r}_{b}$ and $\hat{r}_{s}$ are shown in Figure \ref{fig:rb_rs}. This example assumes each deputy is modeled as a 6U CubeSat in Low Earth Orbit (LEO), where $n=0.001027$ radians per second and $m=12$ kg.

\subsection{Safety Constraints}

While many safety constraints could be developed for this task, the following constraints define $\mathcal{C}_{\rm A}$ for this analysis. The constraints are defined for $N$ deputies, $\forall i \in \mathbb{Z}_{1:N}$, $\forall j \in \mathbb{Z}_{1:N}$, $i\neq j$.

\subsubsection{Safe Separation.}
Each deputy spacecraft shall not collide with the chief. This constraint is defined as,
\begin{equation}
    \varphi_1(\boldsymbol{x}) := \Vert \boldsymbol{p}_i \Vert_2 - (r_{\rm d}+r_{\rm c}) \geq 0,
\end{equation}
where $r_{\rm d}$ is the collision radius of the deputy and $r_{\rm c}$ is the collision radius of the chief. Additionally, each deputy spacecraft shall not collide with another deputy. This constraint is defined as,
\begin{equation}
    \varphi_2(\boldsymbol{x}) := \Vert \boldsymbol{p}_i - \boldsymbol{p}_j \Vert_2 - 2r_{\rm d} \geq 0.
\end{equation}

\subsubsection{Dynamic Speed Constraint.}

The speed of each deputy shall decrease as it moves closer to the chief. This reduces risk of a high speed collision, as well as risk in the event of a fault \cite{mote2021natural}. Additionally, each deputy should be moving slow enough to appropriately inspect the chief. This constraint is defined as,
\begin{equation}
    \varphi_3(\boldsymbol{x}) := \nu_0 + \nu_1\Vert \boldsymbol{p}_i \Vert_2 - \Vert \boldsymbol{v}_i \Vert_2 \geq 0,
\end{equation}
where $\nu_0$ is a minimum allowable docking speed, $\nu_1$ is a constant rate at which $\boldsymbol{p}$ shall decrease, and $\boldsymbol{v}_i=[\dot{x}, \dot{y}, \dot{z}]^T$.

\subsubsection{Keep Out Zone.}

\begin{figure}[htb!]
    \centering
    \includegraphics[width=.5\textwidth]{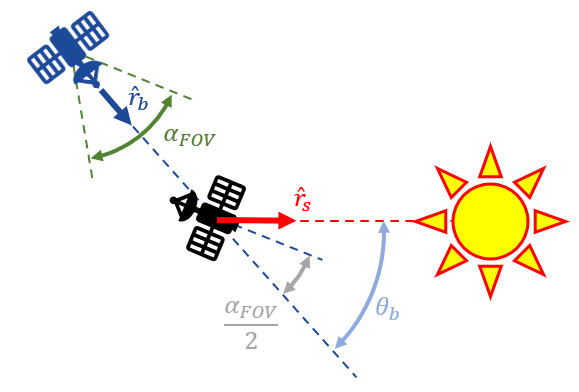}
    \caption{Sun keep out zone.}
    \label{fig:a_fov}
\end{figure}

Each deputy shall not align its sensor with the Sun. This constraint is defined as,
\begin{equation}
    \varphi_4(\boldsymbol{x}) := \theta_{b_i} - \frac{\alpha_{FOV}}{2} \geq 0,
\end{equation}
where $\theta_{b_i}$ is the angle between the deputy's sensor boresight and the Sun, and $\alpha_{FOV}$ is the sensor's field of view, as shown in Figure \ref{fig:a_fov}. $\theta_{b_i}$ is found using the dot product,
\begin{equation}
    \theta_{b_i} = \arccos{(\hat{r}_{b_i} \cdot \hat{r}_s)}.
\end{equation}

In the event that a deputy must change its orientation to inspect another deputy, an additional keep out zone constraint must be developed. In this case, each deputy shall not align its sensor with the Sun if it were to point its sensor at another deputy. This constraint is defined as,
\begin{equation}
    \varphi_5(\boldsymbol{x}) := \theta_{b,d_i} - \frac{\alpha_{FOV}}{2} \geq 0,
\end{equation}
where $\theta_{b,d_i}$ is defined as,
\begin{equation}
    \theta_{b,d_i} = \arccos{\left(\frac{\boldsymbol{p}_i - \boldsymbol{p}_j}{\Vert \boldsymbol{p}_i - \boldsymbol{p}_j \Vert_2} \cdot \hat{r}_s\right)}.
\end{equation}

\subsubsection{Keep In Zone.}
Each deputy shall not travel too far from the chief, such that they remain in a specified proximity. This constraint is defined as,
\begin{equation}
    \varphi_6(\boldsymbol{x}) := r_{\rm max} - \boldsymbol{p}_i \geq 0,
\end{equation}
where $r_{\rm max}$ is the maximum relative distance from the chief.

\subsubsection{Passively Safe Maneuvers.}
Each deputy shall not collide with the chief in the event of a fault or loss of power, where it may not be able to use its thrusters. That is, if $\boldsymbol{u}=0$ for an extended period of time, $\varphi_1(\boldsymbol{x})$ shall be enforced for the entire time period. The closed form solution to the Clohessy-Wiltshire equations can be used to determine $\boldsymbol{p}_i$ for any point in time, where,
\begin{equation}
\begin{gathered}
    x(t) = (4-3\cos{nt})x_0 + \frac{\sin{nt}}{n}\dot{x}_0 + \frac{2}{n}(1-\cos{nt})\dot{y}_0, \\
    y(t) = 6(\sin{nt}-nt)x_0 + y_0 - \frac{2}{n}(1-\cos{nt})\dot{x}_0 + \frac{4\sin{nt}-3nt}{n}\dot{y}_0, \\
    z(t) = z_0\cos{nt} + \frac{\dot{z}_0}{n}\sin{nt}.
\end{gathered}
\end{equation}
This trajectory is known as a Free Flight Trajectory (FFT). Letting $\boldsymbol{p}_i=[x_0, y_0, z_0]^T$ and $\boldsymbol{p}_i(t)=[x(t), y(t), z(t)]^T$, the constraint is defined as,
\begin{equation}
    \varphi_7(\boldsymbol{x}) := \inf_{t \in [t_0, t_0+T]} \Vert \boldsymbol{p}_i(t) \Vert_2 - (r_{\rm d}+r_{\rm c}) \geq 0,
\end{equation}
where $T$ is the time period to evaluate over starting at $t_0$. Note that safety is only guaranteed for all time if $T=\infty$, but for practical implementation $T$ is a finite value.
Additionally, each deputy shall not collide with another deputy during a FFT. This constraint is defined as,
\begin{equation}
    \varphi_8(\boldsymbol{x}) := \inf_{t \in [t_0, t_0+T]} \Vert \boldsymbol{p}_i(t)-\boldsymbol{p}_j(t) \Vert_2 - 2r_{\rm d} \geq 0.
\end{equation}

\subsubsection{Structural Damage Threshold.}

Each deputy shall not maneuver aggressively with high velocities. This also ensures each deputy is moving slow enough to appropriately inspect the chief. This is defined in terms of three separate constraints,
\begin{equation}
\begin{gathered}
    \varphi_9(\boldsymbol{x}) := v_{\rm max}^2 - \dot{x}_i^2\geq 0, \quad \varphi_{10}(\boldsymbol{x}) := v_{\rm max}^2 - \dot{y}_i^2\geq 0, \\ \varphi_{11}(\boldsymbol{x}) := v_{\rm max}^2 - \dot{z}_i^2\geq 0,
\end{gathered}
\end{equation}
where $v_{\rm max}$ is the maximum allowable velocity. In addition, each deputy shall remain within the bounds of its actuation limits. This is enforced through the admissible control set $\mathcal{U}$ such that $\boldsymbol{u}_i \in [-u_{\rm max},u_{\rm max}]^3$.

\subsubsection{Fuel Limit.}
The deputy shall adhere to a maximum cumulative fuel use limit, which is considered in terms of $\Delta v$. This constraint is defined as,
\begin{equation}
    \varphi_{12}(\boldsymbol{x}) := \Delta v_{\rm max} - \Delta v_i \geq 0,
\end{equation}
where $\Delta v_{\rm max}$ is the maximum $\Delta v$ use, either for the mission or for the life of the spacecraft. At a given time $t$, $\Delta v_i$ is defined as,
\begin{equation}
    \Delta v_i = \int_{t_0}^{t} \frac{F_{{\rm total}_i}}{m},
\end{equation}
where $F_{\rm total}=|F_x|+|F_y|+|F_z|$ is the total thrust of the spacecraft.


\subsubsection{Values.}

The values used for this problem are defined in Table \ref{tab:constraint_values}.

\begin{table}[htb!]
    \centering
    \caption{Safety Constraint Values}
    \begin{tabular}{c|c} \hline
        Parameter & Value \\ \hline
        $m$ & 12 kg \\
        $n$ & 0.001027 rad/s \\
        $r_{\rm d}$ & 5 m \\
        $r_{\rm c}$ & 5 m \\
        $\nu_0$ & 0.2 m/s \\
        $\nu_1$ & 2n rad/s \\
        $\Delta v_{\rm max}$ & 20 m/s \\
        $\alpha_{FOV}$ & 60 degrees \\
        $r_{\rm max}$ & 1000 m \\
        $T$ & 500 s \\
        $v_{\rm max}$ & 1 m/s \\
        $u_{\rm max}$ & 1 N \\
        \hline
    \end{tabular}
    \label{tab:constraint_values}
\end{table}

\subsection{Control Invariant Constraints} \label{sec:control_invariant_constraints}

Due to the admissible control set $\mathcal{U}$, the constraints that define $\mathcal{C}_{\rm A}$ are not all suitable to explicitly define safety. Instead, control invariant safety constraints must be developed that define $\mathcal{C}_{\rm S}$. These constraints are defined as follows. Note that the fuel limit is not considered to be part of $\mathcal{C}_{\rm S}$ because fuel use is directly tied to actuation, so limiting fuel use would prevent the RTA from enforcing all other constraints. An alternate solution to enforcing the fuel limit will be discussed in the next section.

\subsubsection{Safe Separation.}
To maintain safe separation, each deputy spacecraft must consider when to start slowing down to avoid a collision \cite{hibbard2022guaranteeing}. First, consider the projection of the deputy's velocity onto its position vector,
\begin{equation}
    \boldsymbol{v}_{{pr}_i} = \frac{\langle \boldsymbol{v}_i, \boldsymbol{p}_i \rangle}{\Vert \boldsymbol{p}_i \Vert_2}.
\end{equation}
To avoid a collision, the deputy must slow $\boldsymbol{v}_{{pr}_i}$ to zero. Therefore, the following constraint must be satisfied,
\begin{equation} \label{eq:w_int}
    \Vert \boldsymbol{p}_i \Vert_2 + \int_{t_0}^{t_0+T_b} \boldsymbol{v}_{{pr}_i} + a_{\rm max}t \, dt \geq r_{\rm d}+r_{\rm c},
\end{equation}
where $a_{\rm max}$ is the maximum acceleration of the deputy and $T_b=(0-\boldsymbol{v}_{{pr}_i})/a_{\rm max}$ is the time for $\boldsymbol{v}_{{pr}_i}$ to reach zero. $a_{\rm max}$ is found by considering the worst-case acceleration due to natural motion from the system dynamics in Eq. \eqref{eq: system dynamics}. This occurs when $\boldsymbol{x}=[-\Vert {\boldsymbol{r}_{\rm H}} \Vert,0,0,-v_{\rm max},-v_{\rm max},-v_{\rm max}]^T$, which results in,
\begin{equation}
    a_{\rm max} = \frac{u_{\rm max}}{m} - 3n^{2}\Vert {\boldsymbol{r}_{\rm H}} \Vert - 2nv_{\rm max}.
\end{equation}
By computing the integral in Eq. \eqref{eq:w_int} and noting that the constraint only needs to be enforced when the deputy is moving towards the chief ($\boldsymbol{v}_{{pr}_i}<0$), this becomes,
\begin{equation}
    h_1(\boldsymbol{x}) := \sqrt{2 a_{\rm max} [\Vert \boldsymbol{p}_i \Vert_2 - (r_{\rm d}+r_{\rm c})]} + \boldsymbol{v}_{{pr}_i} \geq 0.
\end{equation}
Similarly, each deputy must consider when to slow down to avoid collision with another deputy. This constraint is defined as,
\begin{equation}\label{eq:h2}
    h_2(\boldsymbol{x}) := \sqrt{4 a_{\rm max} (\Vert \boldsymbol{p}_i - \boldsymbol{p}_j\Vert_2 - 2r_{\rm d})} +\boldsymbol{v}_{{pr}_{ij}} \geq 0,
\end{equation}
where,
\begin{equation}
    \boldsymbol{v}_{{pr}_{ij}} = \frac{\langle \boldsymbol{v}_i-\boldsymbol{v}_j, \boldsymbol{p}_i-\boldsymbol{p}_j \rangle}{\Vert \boldsymbol{p}_i-\boldsymbol{p}_j \Vert_2} .
\end{equation}
Note that in Eq. \eqref{eq:h2}, $a_{\rm max}$ is multiplied by 4 because both deputies can slow down to avoid a collision.

\subsubsection{Dynamic Speed Constraint.}
From Reference~\citenum{dunlap2021comparing}, it can be shown that $\varphi_3$ is control invariant if $u_{\rm max}$ adheres to the following limit,
\begin{equation} \label{eq:proof1_4}
    \nu_{1} \sqrt{3} v_{\rm max} + 3n^{2} \frac{\sqrt{3} v_{\rm max} - \nu_0}{\nu_1} + 2nv_{\rm max} \leq \frac{u_{\rm max}}{m}.
\end{equation}
This limit ensures that at the point of worst-case acceleration, $u_{\rm max}$ is large enough to overcome the natural motion, thus preventing the deputy's speed from increasing and violating $\varphi_3$. If the constraint holds at this point, then it will hold for all other points. Given the values from Table \ref{tab:constraint_values}, the limit holds and therefore $h_3=\varphi_3$.

\subsubsection{Keep Out Zone.}
To maintain safe separation from the conical keep out zone, the constraint is formulated in terms of position rather than angles \cite{hibbard2022guaranteeing}. In this case, the unit vector of the cone, $\hat{r}_c$, is assumed to align with the $-\hat{r}_s$ vector and rotate in the $\hat{x}-\hat{y}$ plane at the rate $n$. The vector pointing from $\boldsymbol{p}$ to its projection on $\hat{r}_c$ is given by,
\begin{equation}
    \boldsymbol{p}_{\hat{r}_{c}} = \boldsymbol{p} - \langle \boldsymbol{p}, \hat{r}_c \rangle \hat{r}_c.
\end{equation}
The projection of $\boldsymbol{p}$ onto the cone is then,
\begin{equation}
    \boldsymbol{p}_{c} = \boldsymbol{p} + \sin{\theta}\cos{\theta}\left(\Vert \boldsymbol{p}_{\hat{r}_{c}} \Vert_2 \hat{r}_c + \frac{\langle \boldsymbol{p}, \hat{r}_c \rangle \boldsymbol{p}_{\hat{r}_{c}}}{\Vert \boldsymbol{p}_{\hat{r}_{c}} \Vert_2} \right) - \cos^2{\theta} \boldsymbol{p}_{\hat{r}_{c}} - \sin^2{\theta} \langle \boldsymbol{p}, \hat{r}_c \rangle \hat{r}_c,
\end{equation}
where $\theta = \alpha_{FOV}/2$. This defines the closest position on the cone to the deputy. Similar to the safe separation constraint, the deputy must slow down to avoid entering the cone. This can be written as,
\begin{equation}
    h_{KOZ}(\boldsymbol{x}) := \sqrt{2 a_{\rm max} \Vert \boldsymbol{p}- \boldsymbol{p}_{c}\Vert_2} + \boldsymbol{v}_{pr,c} \geq 0,
\end{equation}
where $\boldsymbol{v}_{pr,c}=\langle \boldsymbol{v}-\boldsymbol{v}_{c}, \boldsymbol{p}-\boldsymbol{p}_{c} \rangle / \Vert \boldsymbol{p}-\boldsymbol{p}_{c} \Vert_2$
and $\boldsymbol{v}_{c}=[0, 0, n] \times \boldsymbol{p}_{c}$. 
For the case of each deputy pointing towards the chief, $h_4=h_{KOZ}$ where $\boldsymbol{p}=\boldsymbol{p}_i$ and $\boldsymbol{v}=\boldsymbol{v}_i$. For the multi-agent keep out zone, $h_5=h_{KOZ}$ where $\boldsymbol{p}=\boldsymbol{p}_i-\boldsymbol{p}_j$ and $\boldsymbol{v}=\boldsymbol{v}_i-\boldsymbol{v}_j$ when $\theta_{b,d_i} > \frac{\pi}{2}$, and otherwise $\boldsymbol{p}=\boldsymbol{p}_j-\boldsymbol{p}_i$ and $\boldsymbol{v}=\boldsymbol{v}_j-\boldsymbol{v}_i$. This sign change allows each deputy to recognize when one is pointing towards the Sun.

\subsubsection{Keep In Zone.}
Again similar to the safe separation constraint, the deputy must slow down to avoid reaching $r_{\rm max}$ \cite{hibbard2022guaranteeing}. This can be written as,
\begin{equation}
    h_6(\boldsymbol{x}) := \sqrt{2 a_{\rm max} (r_{\rm max} - \Vert \boldsymbol{p}_i \Vert_2)} - \boldsymbol{v}_{{pr}_i} \geq 0.
\end{equation}

\subsubsection{Passively Safe Maneuvers.}
This constraint was constructed in a way that already considers the trajectory of the deputy spacecraft. Because of this, safety is guaranteed for all time if $T=\infty$, and therefore $h_7=\varphi_7$ and $h_8=\varphi_8$.

\subsubsection{Structural Damage Threshold.}
From Reference~\citenum{dunlap2021comparing}, it can be shown that $\varphi_9$, $\varphi_{10}$, and $\varphi_{11}$ are respectively control invariant if $u_{\rm max}$ adheres to the following limits,
\begin{equation} \label{eq:proof2}
    3n^2r_{\rm max} + 2nv_{\rm max} \leq \frac{u_{\rm max}}{m}, \quad 2nv_{\rm max} \leq \frac{u_{\rm max}}{m}, \quad n^2r_{\rm max} \leq \frac{u_{\rm max}}{m}.
\end{equation}
These limits again ensure that at the point of worst-case acceleration, $u_{\rm max}$ is large enough to overcome the natural motion, thus preventing the deputy's speed from increasing and violating the constraints. If the constraints hold at this point, then they will hold for all other points. Given the values from Table \ref{tab:constraint_values}, the limits hold and therefore $h_9=\varphi_9$, $h_{10}=\varphi_{10}$, and $h_{11}=\varphi_{11}$.

\subsection{Centralized vs. Decentralized RTA}

The control invariant constraints defined in the previous section can be enforced using ASIF in two different ways. First, the control of all deputies can be decentralized, where each deputy controls itself and uses a separate RTA filter. Each deputy has knowledge of the state of all other deputies, but they do not have knowledge of the control of all others. Decentralized RTA is beneficial because it allows each agent to operate independently. Second, the control of all deputies can be centralized, where all deputies are controlled by a central controller and there is then only one RTA filter. This central controller has knowledge of the state and control of all deputies. Centralized RTA is beneficial because it allows all constraints for all agents to be considered and enforced at the same time.

\subsection{Switching-Based Fuel Limit} \label{sec:fuel_lim}

If a deputy violates the fuel limit, the RTA should guide the deputy spacecraft to a safe state where no fuel can be used. In this case, closed elliptical Natural Motion Trajectories (eNMT) \cite{frey2017constrained} are used because they are ``parking orbits" where the deputy can orbit the chief without using any fuel. Closed eNMTs centered at the origin of Hill's frame satisfy the following,
\begin{equation}
    \Dot{y}(0)=-2nx(0), \quad \dot{x}(0)= \frac{n}{2}y(0).
\end{equation}

While not all of the constraints in the previous section can be enforced on an eNMT, the deputy will at minimum be able to maintain safe separation from the chief.
To enforce the fuel limit, the RTA uses a switching-based approach that switches control to a backup controller when the constraint is violated. This is sometimes referred to as a \textit{latched} RTA approach, as the system does not switch back and forth between the primary and backup controllers, but rather remains latched to the backup controller until a specified condition is met. The switching filter used for this analysis can be defined as follows.

\noindent \rule{1\columnwidth}{0.7pt}
\noindent \textbf{Switching Filter}
\begin{equation}
\begin{array}{rl}
\boldsymbol{u}_{\rm act}(\boldsymbol{x})=
\begin{cases} 
\boldsymbol{u}_{\rm des}(\boldsymbol{x}) & {\rm if}\quad \varphi_{12}(\phi^{\boldsymbol{u}_{\rm b}}(t, \boldsymbol{x})) \geq 0, \quad \forall t \in [t_0, t_0+T]  \\ 
\boldsymbol{u}_{\rm b}(\boldsymbol{x})  & {\rm if}\quad otherwise
\end{cases}
\end{array}\label{eq:switching}
\end{equation}
\noindent \rule[7pt]{1\columnwidth}{0.7pt}

Here, $\phi^{\boldsymbol{u}_{\rm b}}$ represents a prediction of the state $\boldsymbol{x}$ at time $t$ under the backup control law $\boldsymbol{u}_{\rm b}$. Note that for practical implementation, this trajectory is only simulated for a finite time period $T$. The backup controller used for this analysis is an LQR tracking controller that guides the deputy to the nearest eNMT. The backup control law is,
\begin{equation}
    \boldsymbol{u}_{\rm b} = -K_1 \boldsymbol{e} (\boldsymbol{x}-\boldsymbol{x}_{\rm des}) - K_2 \boldsymbol{z},
\end{equation}
where $K_1$ and $K_2$ are LQR gains, $\boldsymbol{e}=\boldsymbol{x}-\boldsymbol{x}_{\rm des}$ is the error between the current and desired states, $\boldsymbol{z}$ tracks the integral of the error over time, and $\boldsymbol{x}_{\rm des}$ is defined as,
\begin{equation}
    \boldsymbol{x}_{\rm des} = [x, y, z, ny/2, -2nx, \dot{z}]^T.
\end{equation}

\noindent Note that a traditional LQR controller cannot be used because $\boldsymbol{x}_{\rm des}$ is not a stationary point.

\section{Results}

This section presents and discusses simulation results of the centralized and decentralized ASIF RTA, Monte Carlo simulations, and simulation results of the switching-based fuel limit RTA.

\subsection{Simulation Results} \label{sec:sim_results}

\begin{figure}[htb!]
\centering
\begin{subfigure}[t]{0.45\textwidth}
    \begin{center}
    \includegraphics[width=\textwidth]{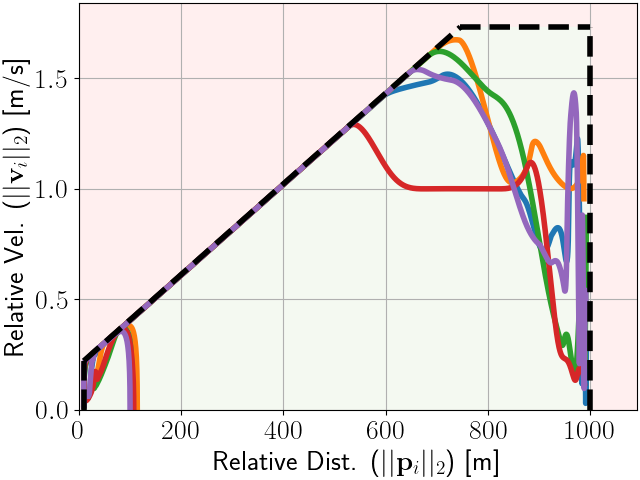}
    \captionsetup{width=.9\linewidth}
    \caption{Chief safe separation, dynamic speed, keep in zone, and velocity limit constraints. Colored lines represent each deputy.}
    \label{fig:rel_vel}
    \end{center}
\end{subfigure}
\begin{subfigure}[t]{0.45\textwidth}
    \begin{center}
    \includegraphics[width=\textwidth]{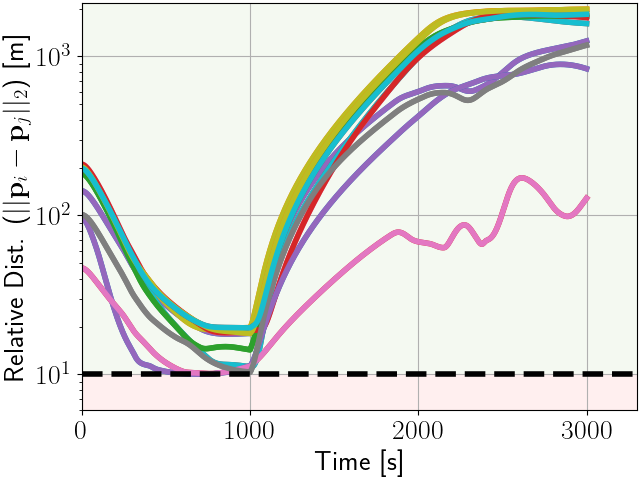}
    \captionsetup{width=.9\linewidth}
    \caption{Deputy safe separation constraint. Colored lines represent the interaction between two deputies.}
    \label{fig:pos_deputy}
    \end{center}
\end{subfigure}
\begin{subfigure}[t]{0.45\textwidth}
    \begin{center}
    \includegraphics[width=\textwidth]{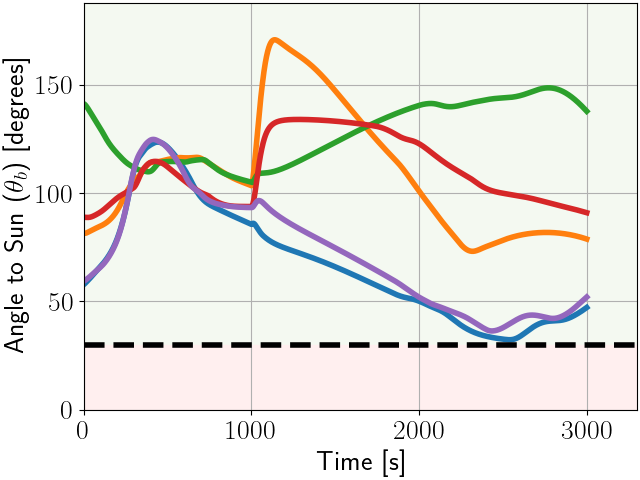}
    \captionsetup{width=.9\linewidth}
    \caption{Keep out zone constraint. Colored lines represent each deputy.}
    \label{fig:chief_sun}
    \end{center}
\end{subfigure}
\begin{subfigure}[t]{0.45\textwidth}
    \begin{center}
    \includegraphics[width=\textwidth]{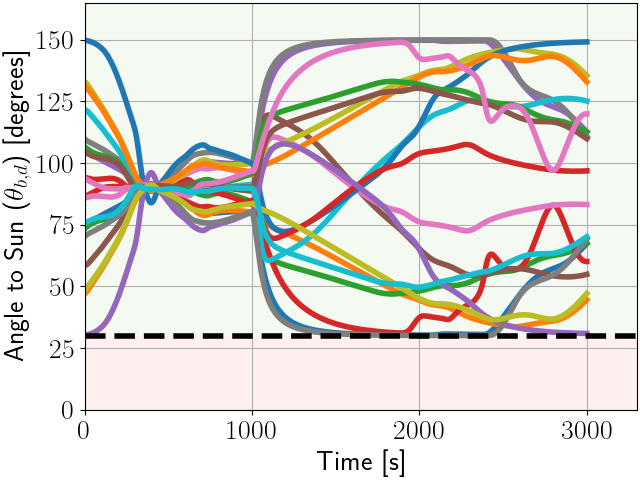}
    \captionsetup{width=.9\linewidth}
    \caption{Multi-agent keep out zone constraint. Colored lines represent the interaction between two deputies.}
    \label{fig:deputy_sun}
    \end{center}
\end{subfigure}
\begin{subfigure}[t]{0.45\textwidth}
    \begin{center}
    \includegraphics[width=\textwidth]{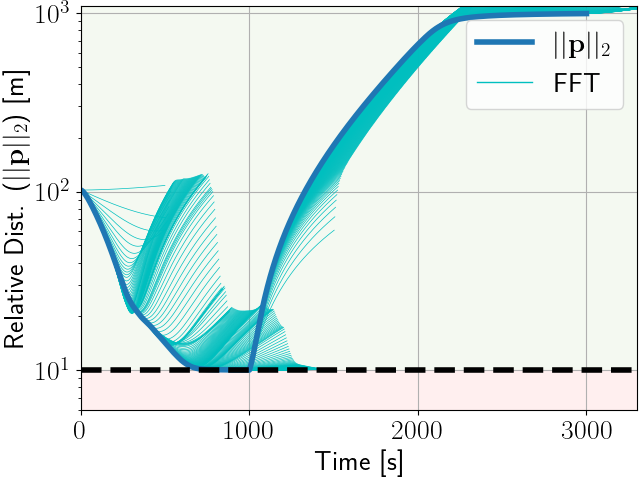}
    \captionsetup{width=.9\linewidth}
    \caption{Passively safe maneuvers constraint for one deputy. The blue line represents the actual trajectory, while the cyan lines represent potential FFTs.}
    \label{fig:PSM}
    \end{center}
\end{subfigure}
\begin{subfigure}[t]{0.45\textwidth}
    \begin{center}
    \includegraphics[width=\textwidth]{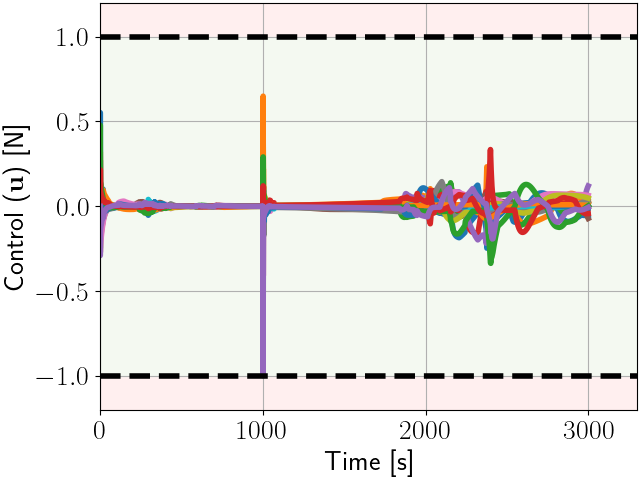}
    \captionsetup{width=.9\linewidth}
    \caption{Control limits. Colored lines represent components of the control vector for each deputy.}
    \label{fig:control}
    \end{center}
\end{subfigure}
\caption{Simulation results for centralized ASIF RTA.}
\label{fig:Sim}
\end{figure}

To evaluate the ability of the RTA to enforce all constraints, a simulation of the inspection task with centralized RTA is shown in Figure \ref{fig:Sim}. In this figure, safe regions are shaded green and unsafe regions are shaded red. The primary controller for this simulation is an aggressive LQR controller designed to violate the safety constraints. Note that the purpose of this primary controller is not to complete the inspection task, but rather to show that RTA assures safety of all constraints. The simulation is run with 5 deputies for 3,000 seconds, where for the first 1,000 seconds the primary controller attempts to move each deputy to the origin, and for the last 2,000 seconds the primary controller attempts to move each deputy to a distance greater than $r_{\rm max}$. Figure \ref{fig:Sim} shows that the centralized RTA assures safety of each constraint simultaneously.

\begin{figure}[htb!]
\centering
\begin{subfigure}[t]{0.45\textwidth}
    \begin{center}
    \includegraphics[width=\textwidth]{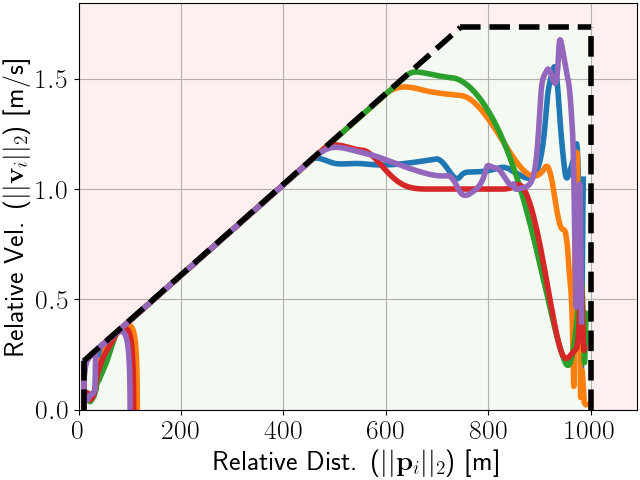}
    \captionsetup{width=.9\linewidth}
    \caption{Chief safe separation, dynamic speed, keep in zone, and velocity limit constraints. Colored lines represent each deputy.}
    \label{fig:rel_vel_dec}
    \end{center}
\end{subfigure}
\begin{subfigure}[t]{0.45\textwidth}
    \begin{center}
    \includegraphics[width=\textwidth]{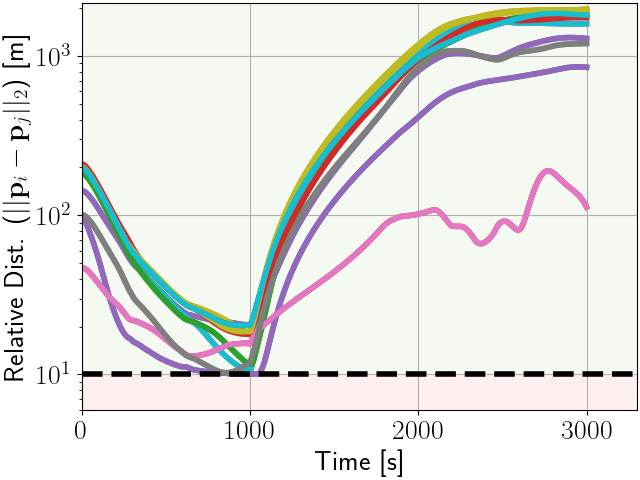}
    \captionsetup{width=.9\linewidth}
    \caption{Deputy safe separation constraint. Colored lines represent the interaction between two deputies.}
    \label{fig:pos_deputy_dec}
    \end{center}
\end{subfigure}
\begin{subfigure}[t]{0.45\textwidth}
    \begin{center}
    \includegraphics[width=\textwidth]{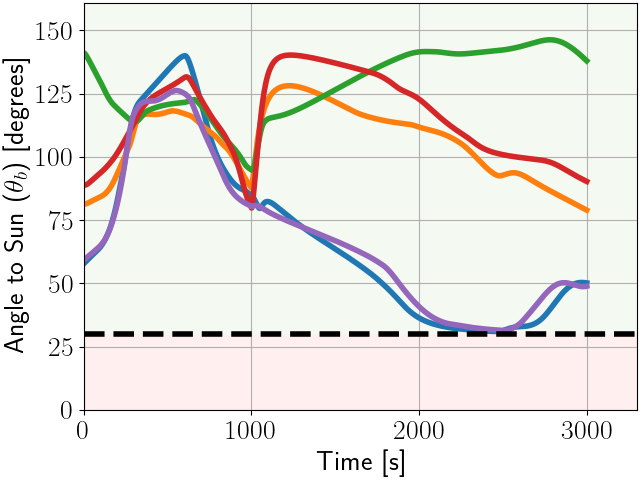}
    \captionsetup{width=.9\linewidth}
    \caption{Keep out zone constraint. Colored lines represent each deputy.}
    \label{fig:chief_sun_dec}
    \end{center}
\end{subfigure}
\begin{subfigure}[t]{0.45\textwidth}
    \begin{center}
    \includegraphics[width=\textwidth]{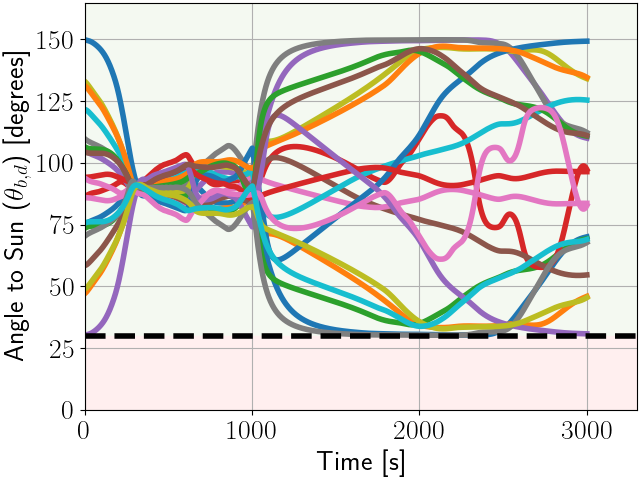}
    \captionsetup{width=.9\linewidth}
    \caption{Multi-agent keep out zone constraint. Colored lines represent the interaction between two deputies.}
    \label{fig:deputy_sun_dec}
    \end{center}
\end{subfigure}
\begin{subfigure}[t]{0.45\textwidth}
    \begin{center}
    \includegraphics[width=\textwidth]{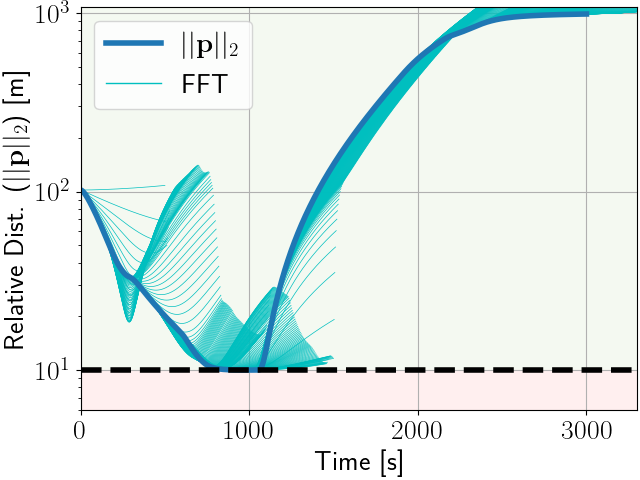}
    \captionsetup{width=.9\linewidth}
    \caption{Passively safe maneuvers constraint for one deputy. The blue line represents the actual trajectory, while the cyan lines represent potential FFTs.}
    \label{fig:PSM_dec}
    \end{center}
\end{subfigure}
\begin{subfigure}[t]{0.45\textwidth}
    \begin{center}
    \includegraphics[width=\textwidth]{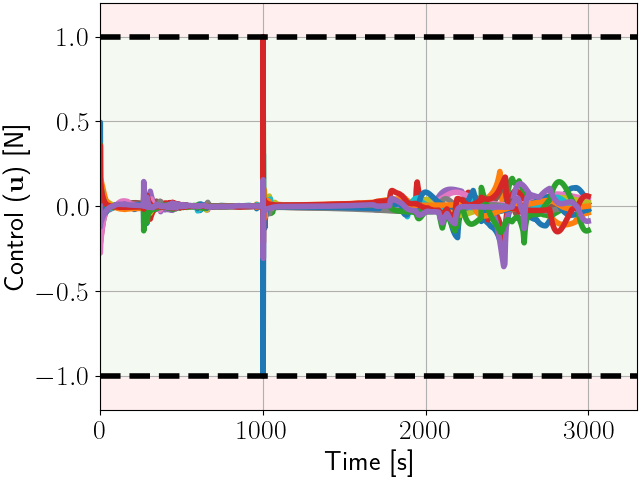}
    \captionsetup{width=.9\linewidth}
    \caption{Control limits. Colored lines represent components of the control vector for each deputy.}
    \label{fig:control_dec}
    \end{center}
\end{subfigure}
\caption{Simulation results for decentralized ASIF RTA.}
\label{fig:Sim_dec}
\end{figure}

The same simulation was run again with decentralized RTA, where the results are shown in Figure \ref{fig:Sim_dec}. In this case, the decentralized RTA is also able to assure safety of each constraint simultaneously using a slightly different behavior.

\subsection{Monte Carlo}

In order for human operators to trust an RTA filter, it must be shown to assure safety for all possible scenarios. One way to achieve this is through the use of Monte Carlo analysis, which runs a large number of simulations to cover a substantial amount of the state space. While not every state can be tested, this does allow the designer to verify that RTA can assure safety for almost any state. For this analysis, 2,000 simulations with 5 deputies were run for 500 seconds each. No primary control input was used during these simulations. Latin hypercube sampling was used to compute initial conditions, where initial conditions that violated any safety constraint were removed and resampled. The same set of 2,000 initial conditions was used to simulate the centralized and decentralized RTA filters.

\begin{figure}[htb!]
\centering
\begin{subfigure}[t]{0.4\textwidth}
    \begin{center}
    \includegraphics[width=\textwidth]{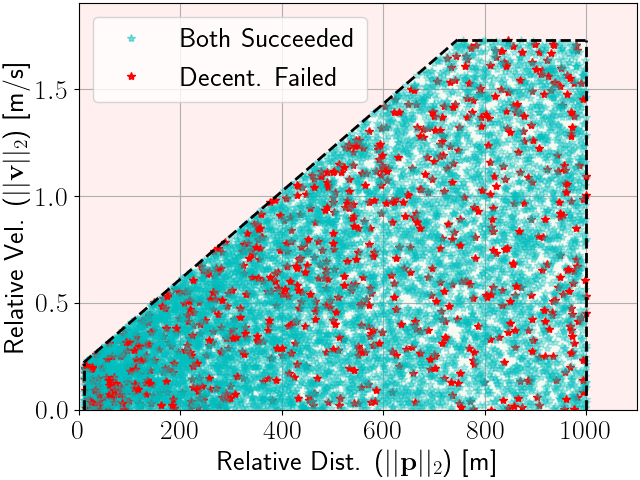}
    \captionsetup{width=.9\linewidth}
    \caption{Chief safe separation, dynamic speed, keep in zone, and velocity limit constraints.}
    \label{fig:MCpos}
    \end{center}
\end{subfigure}
\begin{subfigure}[t]{0.4\textwidth}
    \begin{center}
    \includegraphics[width=\textwidth]{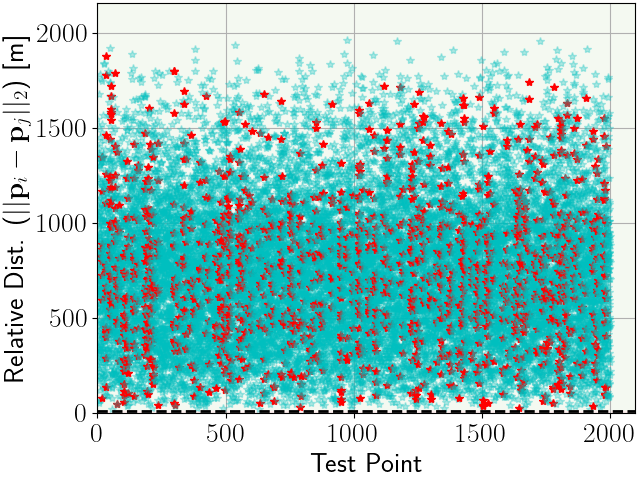}
    \captionsetup{width=.9\linewidth}
    \caption{Deputy safe separation constraint.}
    \label{fig:MCdeppos}
    \end{center}
\end{subfigure}
\begin{subfigure}[t]{0.4\textwidth}
    \begin{center}
    \includegraphics[width=\textwidth]{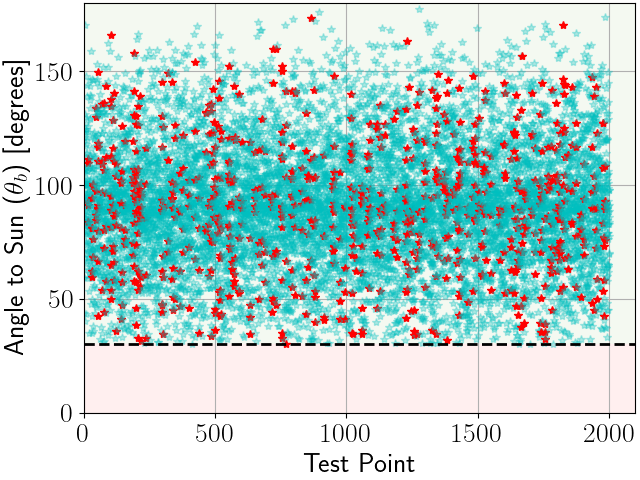}
    \captionsetup{width=.9\linewidth}
    \caption{Keep out zone constraint.}
    \label{fig:MCsun}
    \end{center}
\end{subfigure}
\begin{subfigure}[t]{0.4\textwidth}
    \begin{center}
    \includegraphics[width=\textwidth]{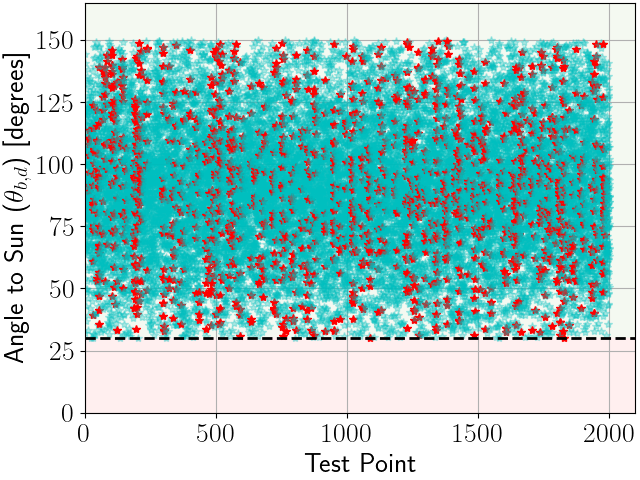}
    \captionsetup{width=.9\linewidth}
    \caption{Multi-agent keep out zone constraint.}
    \label{fig:MCdepsun}
    \end{center}
\end{subfigure}
\caption{Monte Carlo simulation results. Each star represents the initial conditions of a test case, where all 2,000 test cases are shown in each figure. 100\% of all centralized RTA test cases were successful. 90.95\% of decentralized RTA test cases were successful (where both succeeded, shown in cyan), and 9.05\% failed (shown in red).}
\label{fig:MonteCarlo}
\end{figure}

Figure \ref{fig:MonteCarlo} shows the results of the Monte Carlo simulations for both centralized and decentralized RTA, where failure refers to any initial condition where RTA failed to assure safety for at least one deputy or the QP failed to find a safe control input that satisfied all constraints. Overall, the centralized RTA was able to assure safety of all constraints for 100\% of the points, while the decentralized RTA was able to assure safety of all constraints for 90.95\% of the points.

The points where the decentralized RTA failed were all due to the multi-agent keep out zone, where each deputy shall not align its sensor with the Sun if it were to point its sensor at another deputy. These failures were due to the design of the constraint, where each deputy is assumed to accelerate at the maximum rate to avoid entering the exclusion zone. In the event that more than one deputy is attempting to adhere to conflicting constraints at the same time, it may not be able to accelerate at the maximum rate. For the case of decentralized RTA, the other deputies are not aware of the conflicting constraints, causing the RTA to eventually fail. For the case of centralized RTA, the centralized controller is aware of all constraints, and can adjust the control of each deputy to avoid this scenario. Multiple failure mitigation strategies can be used for decentralized RTA, including adjusting the strengthening function $\alpha(x)$ or removing/relaxing the multi-agent keep out zone constraint when necessary. The Monte Carlo simulation was run again using the same 2,000 initial conditions, but removing the multi-agent keep out zone, and the decentralized RTA was then able to assure safety of all constraints for 100\% of the points.

\subsection{Switching-Based Fuel Limit}

\begin{figure}[htb!]
\centering
\begin{subfigure}[t]{0.5\textwidth}
    \begin{center}
    \includegraphics[width=\textwidth]{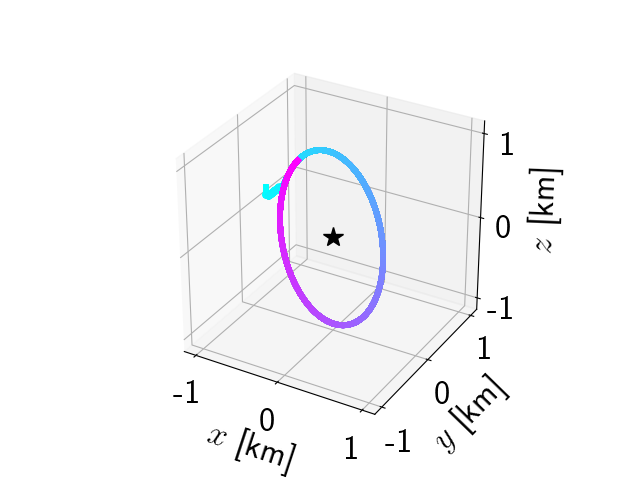}
    \caption{The trajectory of the deputy switches from an aggressive LQR controller to a backup controller maneuvering the deputy to an eNMT, transitioning from cyan to magenta as time increases.}
    \label{fig:fuel_traj}
    \end{center}
\end{subfigure}
\begin{subfigure}[t]{0.4\textwidth}
    \begin{center}
    \includegraphics[width=\textwidth]{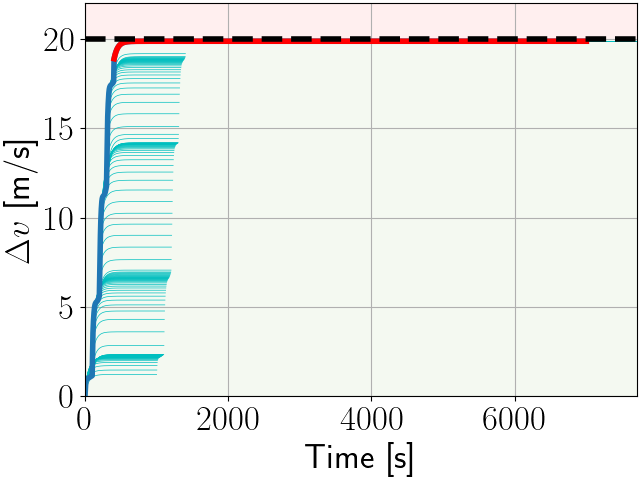}
    \captionsetup{width=.9\linewidth}
    \caption{The RTA switches from an aggressive LQR controller to a backup controller maneuvering the deputy to an eNMT when the line changes from blue to red. Potential trajectories under the backup control law are shown in cyan.}
    \label{fig:fuel}
    \end{center}
\end{subfigure}
\begin{subfigure}[t]{0.4\textwidth}
    \begin{center}
    \includegraphics[width=\textwidth]{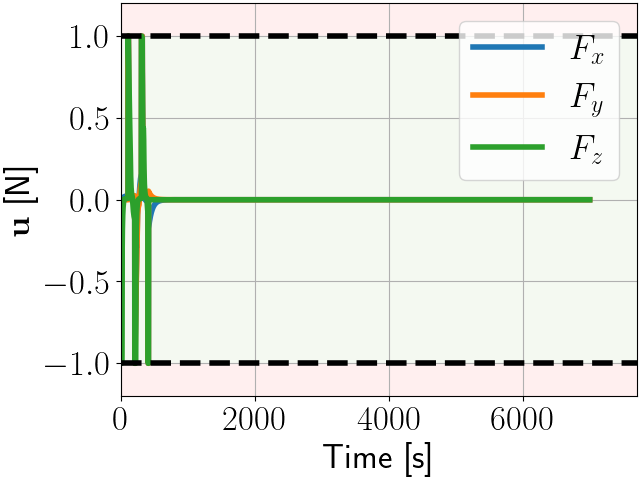}
    \captionsetup{width=.9\linewidth}
    \caption{Control during the simulation, where the force produced in each direction goes to zero once the deputy reaches the eNMT.}
    \label{fig:fuel_control}
    \end{center}
\end{subfigure}
\caption{Simulation results with the switching-based fuel limit.}
\label{fig:FuelSim}
\end{figure}

To evaluate the ability of the switching-based RTA to enforce the fuel limit, another simulation of the inspection task is shown in Figure \ref{fig:FuelSim} with only one deputy. The primary controller for this simulation is an aggressive LQR controller designed to use as much fuel as possible. The simulation is run for 7,000 seconds.

The trajectory of the deputy is shown to maneuver to an eNMT around the chief before the fuel limit is violated in Figure \ref{fig:fuel_traj}, where the position of the deputy changes from cyan to magenta as time increases. Figure \ref{fig:fuel} shows the fuel use constraint, and Figure \ref{fig:fuel_control} shows the control limits. It can be seen that before the total $\Delta v$ exceeds the limit, the system switches to the backup controller, which guides the deputy to the nearest eNMT, where the deputy orbits the chief and $\Delta v$ does not increase for the remainder of the simulation.

\section{Conclusion}

This paper developed several safety constraints for an autonomous spacecraft inspection problem. Most of the constraints were enforced using an ASIF RTA approach, which is an optimization-based approach that is minimally invasive with respect to the primary controller, where individual simulations were used to show that the ASIF RTA is able to enforce all constraints simultaneously. A comparison was made between centralized and decentralized control, where Monte Carlo analysis was used to show that centralized RTA was able to assure safety for all test points, and decentralized RTA was able to assure safety for most test points. Failure mitigation strategies were also presented, including modifying the constraint strengthening function and removing low priority constraints.
While ASIF is a beneficial approach that is able to enforce most constraints, a scenario was also presented where a switching-based RTA approach more effectively enforced a fuel constraint. This analysis shows that as the system and constraint complexity increases, RTA methods can be combined to achieve the best safety assurance technique.


\bibliographystyle{AAS_publication}   
\bibliography{references}

\end{document}